\documentclass[12pt]{article}

\usepackage{amsmath}
\usepackage{amssymb}
\usepackage{amsfonts}
\usepackage{graphicx}
\usepackage{slashed} 
 
\numberwithin{equation}{section}

\begin{document}

\begin{center}
{\large \bf{ The $H_T$ Higgs boson at  the LHC Run 2}}
\end{center}

\vspace*{2cm}

\begin{center}
{
Paolo Cea~\protect\footnote{Electronic address:
{\tt Paolo.Cea@ba.infn.it}}  \\[0.5cm]
{\em INFN - Sezione di Bari, Via Amendola 173 - 70126 Bari,
Italy} }
\end{center}

\vspace*{1.5cm}

\begin{abstract}
\noindent 
We further elaborate on  the proposal that  the  Higgs boson should be a broad heavy resonance, referred to as  true Higgs $H_T$, 
 with mass around  $750 \, GeV$.  
 We stress once again that within the Standard Model  the true Higgs is the  unique possibility to  implement the  spontaneous 
 symmetry breaking of the local gauge symmetry by elementary, relativistic and strictly  local scalar fields.  
We  discuss  the most relevant decay modes  of the $H_T$ boson and estimate they partial decay widths and branching ratios. We  discuss
briefly the experimental signatures of the  $H_T$ Higgs boson and compare with the recent available LHC data  at $\sqrt{s} =13 \,TeV$.
We find that the coupling of the  $H_T$ Higgs boson to fermions is strongly suppressed.
 We also compare our theoretical expectations in the so-called golden channel to the data collected by the ATLAS and CMS Collaborations
at  $\sqrt{s} =13 \,TeV$  with an integrated luminosity of $36.1 \, fb^{-1}$ and $35.9 \, fb^{-1}$ respectively.
We find that  our theoretical expectations are in fair good agreement with the  experimental observations.
Combining the data from both the  LHC Collaborations we obtain an evidence of the heavy Higgs boson  in this channel 
with an estimated statistical significance  of more than three  standard deviations. Finally, we argue that, if the signal
is real, by the end of the  Run 2 both the LHC experiments   will reach in the golden channel a statistical significance of about
 five  standard deviations.
\end{abstract}

\vspace*{1.5cm}
\noindent
The author  declares  that there is no conflict of interest regarding the publication of this paper.

\newpage
\noindent
\section{Introduction}
\label{s-1}
A fundamental feature of the Standard Model is the mechanism of spontaneous symmetry breaking, the so-called 
Brout-Englert-Higgs (BEH) mechanism~\cite{Englert:1964,Higgs:1964,Guralnik:1964,Higgs:1966}.
The first runs of proton-proton collisions at the CERN Large Hadron Collider (LHC) with center-of-mass
energies $\sqrt{s} = 7 \, ,\, 8 \; TeV$ (Run 1) has brought the confirmation of the existence of 
a  boson, named H, which resembles the one which breaks the electroweak symmetry in the Standard Model
 of particle physics~\cite{Aad:2012,Chatrchyan:2012}. 
The combined ATLAS and CMS experiments best estimate of the mass of the H boson was
$m_H  = 125.09 \,  \pm \, 0.24 \; GeV$~\cite{ATLAS-CMS:2015}. Moreover, the results from both LHC experiments, 
as summarized in Refs.~\cite{CMS:2015,ATLAS:2016,ATLAS-CMS:2016},
showed that all measurements of the properties of the new H resonance were consistent with those expected for the
Standard Model Higgs boson. 
Actually, in the LHC Run 1 the strongest signal significance has been obtained from the
decays of the H boson into two vector bosons, $H \rightarrow V V$ where $V = \gamma, W, Z$. In fact, in these channels
the observed signal significance were above five standard deviations~\cite{CMS:2015,ATLAS:2016,ATLAS-CMS:2016}. 
Nevertheless, soon after the evidence of the LHC resonance at  $125 \; GeV$, we already
proposed~\cite{Cea:2012b} that  the $H$ resonance  could be interpreted as a pseudoscalar meson. 
In particular, we showed~\cite{Cea:2016}  that  our pseudoscalar meson  could  mimic the decays of the Standard Model  Higgs boson
 in the  vector  boson decay channels, while  the decays into fermions were strongly  suppressed. 
 Moreover, the main decay channels of this pseudoscalar meson were the hadronic decays that could mimic the decays of
 a putative Higgs boson in the hadronic channels.
We feel that the only save way to distinguish experimentally our pseudoscalar meson from the Standard Model Higgs boson is 
to determine the $CP$ assignment of the H resonance.
 The spin and $CP$ properties of the $H$ boson can be determined by studying
 the tensor structure of its interactions with the electroweak gauge bosons.  The Run 1 experimental analyses
 relied on  discriminant observables chosen to be sensitive to the spin and parity of the signal. 
 In this way it was possible to compare the Standard Model
hypothesis $J^{PC} = 0^{++}$ to  several alternative spin and parity models. 
It turned out that all tested alternative models were excluded with a 
statistical significance of about three standard deviations~\cite{ATLAS:2015d,CMS:2015b}.  In particular, spin-one and spin-two
hypotheses were excluded at a 99 \% CL or higher.  Given the exclusion of the spin-one and spin-two scenarios, constraints were
set on the anomalous couplings of the $H$ resonance to vector bosons by assuming a spin-zero state.
Under the hypothesis that the new resonance is a spin-zero boson, the tensor structure of the interactions of the $H$ boson 
 with two vector bosons  were investigated and limits on CP-odd anomalous contributions were set. 
As a result, in the LHC Run 1 the pseudoscalar hypothesis was excluded at  99 \% CL  from CMS~\cite{CMS:2015b}
and at  97.8  \% CL  from ATLAS ~\cite{ATLAS:2015d}.  However, this conclusion was not strengthened by the recent data from Run 2.
As a matter of fact,  the CMS Collaboration performed the study of the anomalous interactions of the $H$ resonance by using
the full dataset recorded during the Run 2 corresponding to an integrated luminosity of $35.9 \; fb^{-1}$ at $\sqrt{s} = 13 \, TeV$.
Even tough  the number of analyzed $H$ boson in Run 2 was about three times larger than in Run 1, 
the data were consistent with an almost equal mixture  of scalar and pseudoscalar couplings~\cite{CMS:2017d}. 
Similar conclusions have been reached also by the ATLAS Collaboration~\cite{ATLAS:2017b} after collecting
$36.1 \; fb^{-1}$   in the Run 2. 
As a consequence, we may affirm  that  our pseudoscalar interpretation of the $H$ resonance cannot be yet completely excluded. 
Aside from these phenomenological considerations,  we believe that  there are  different and compelling theoretical  motivations 
to doubt on the identification of the $H$ resonance with the Standard Model Higgs boson. 
In fact,  stemming from the known triviality problem, i.e. vanishing of the self-coupling, 
 that affects self-interacting local scalar quantum fields in four space-time dimensions~\cite{Fernandez:1992}, 
 it was evidenced that the Higgs boson condensation
triggering the spontaneous breaking of the local gauge symmetries needs to be dealt with non perturbatively. 
If this is the case, from one hand there is no stability problem for the condensate ground state,  on the other hand the Higgs mass is finitely
related to the vacuum  expectation value of the quantum scalar field and it can be evaluated from first principles.
 Precise non-perturbative numerical simulations indicated that the true Higgs boson, henceforth denoted as
$H_T$, is a  rather heavy resonance with mass around $750 \, GeV$~\cite{Cea:2012}. \\
The aim of the present paper is to  elaborate on the phenomenological consequences of the massive Higgs boson
proposal.  In particular, we will discuss the couplings of the $H_T$ Higgs boson to the massive vector bosons and to fermions,
 the expected production mechanism, and the main decay modes.  
 We organize the paper as following.  In Sect.~\ref{s-2}, for sake of completeness, we briefly illustrate how spontaneous symmetry
 breaking arises in field theories involving scalar fields without quartic self-couplings. Section~\ref{s-3} is devoted to the discussion
 of the couplings of our massive Higgs boson proposal to Standard Model gauge fields. We determine the main decay channels
 of the $H_T$ Higgs boson and we critically examine the couplings to fermions. We also illustrate the main production mechanisms of
 the $H_T$ Higgs boson and estimate the production cross sections at the proton-proton collider at center-of-mass energy
 $\sqrt{s} =13 \,TeV$. In Sect.~\ref{s-4}  we compare  our proposal with available  LHC Run 2 data from   both
 ATLAS and CMS Collaborations.  In Sect.~\ref{s-4-1} we try a quantitative comparison in the so-called golden channel
 of our theory with the recent data collected by the ATLAS and CMS Collaborations at   $\sqrt{s} =13 \,TeV$ corresponding to an
 integrated luminosity of  $36.1 \, fb^{-1}$ and $35.9 \, fb^{-1}$ respectively. Finally,  our concluding remarks are relegated to Sect.~\ref{s-5}. 
\section{Triviality and spontaneous symmetry breaking}
\label{s-2}
Usually the spontaneous symmetry breaking in the Standard Model is implemented
within the perturbation theory which leads to predict that the Higgs boson mass squared  is proportional to $\lambda \, v^2$,
where $\lambda$ is the renormalized scalar self-coupling and  $v \simeq 246 \; GeV$ is the known weak scale. 
 On the other hand, it is known that,  within the non-perturbative  description of spontaneous symmetry breaking in the Standard Model,
 self-interacting scalar fields are subject to the triviality problem~\cite{Fernandez:1992}, namely the renormalized self-coupling
$\lambda \rightarrow  0$ when the ultraviolet cutoff  is sent to infinity.  Strictly speaking, there are  no rigorous proof of triviality.
Nevertheless, there exist several numerical studies  which leave little doubt on the triviality conjecture.  As a consequence, 
within the perturbative approach,  the scalar sector of the Standard Model represents just an effective description  valid only up to some
cut-off scale. If the renormalized self-coupling of the scalar fields vanishes, then one faces with the problem of  the spontaneous symmetry 
breaking mechanism and the related scalar Higgs boson. In fact, naively, one expects that the spontaneous symmetry breaking
mechanism cannot be implemented without the scalar self-coupling  $\lambda$. However,  in Ref.~\cite{Cea:2012},  by means 
of nonperturbative numerical simulations  of the $\lambda \Phi^4$ theory on the lattice,  it was  enlightened  the scenario where
 the Higgs boson without self-interaction  could  coexist with spontaneous symmetry breaking. Moreover,  due to the peculiar rescaling of  
 the Higgs condensate, the relation between the Higgs mass  and $v$ is not the  same as in perturbation theory.  
In fact, remarkably, it turned out that the Higgs mass  were finitely related to $v$.  \\ 
For reader convenience, in the present Section,  following  Ref.~\cite{Cea:2012}, we shall illustrate how  spontaneous symmetry breaking 
could be compatible with triviality. To this end, we consider the simplest scalar field theory, namely a massless real scalar field 
$\Phi$ with quartic self-interaction $ \lambda$ in four space-time dimensions:
\begin{equation} 
\label{2.1} 
{\cal{L}} =\frac{1}{2}(\partial_\mu \Phi_0)^2
- \frac{1}{4} \lambda_0 \ \, \Phi_0^4 \; ,
\end{equation}
where $\lambda_0$ and $\Phi_0$ are the bare coupling and field respectively. As it is well known~\cite{Coleman:1973,Jackiw:1974}, 
in the one-loop approximation the effective potential is given by:
\begin{equation} 
\label{2.2} 
V_{eff}^{1-loop}(\phi_0)   =  \frac{1}{4} \lambda_0 \, \phi_0^4  + \frac{1}{2}  \int \frac{d^3 k}{(2 \pi)^3}  \; 
 \sqrt{  \vec{k}^2  + 3 \lambda_0  \phi_0^2  } \, .
\end{equation}
This last equation shows that the one-loop effective potential is given the vacuum energy of the shifted field in the quadratic approximation.
In fact, let us write:
\begin{equation} 
\label{2.3} 
 \Phi_0 \; = \; \phi_0 \; + \eta \; 
\end{equation}
where $\phi_0$ is the bare uniform scalar condensate, then in this approximation 
the Hamiltonian of the fluctuation $\eta$ over the background $\phi_0$ is:
\begin{equation} 
\label{2.4} 
{\cal{\hat{H}}}_0  = \frac{1}{2}  (\Pi_{\eta} )^2 \; + \; \frac{1}{2}  (\vec{\nabla} \eta)^2 \; + \; \frac{1}{2}  \; ( 3 \lambda_0  \phi_0^2 ) \; \eta^2 \; 
 + \; \frac{1}{4} \lambda_0 \, \phi_0^4 \; .
\end{equation}
After canonical quantization of the quadratic Hamiltonian  ${\cal{\hat{H}}}_0 $ one readily finds that the energy density of the
quantum vacuum in presence of the condensate   $\phi_0$ is given by  Eq.~(\ref{2.2}).
It is, now, easy to see that the one-loop effective potential Eq.~(\ref{2.2}) displays a non-trivial minimum  implying
spontaneous symmetry breaking. However, the minimum of the effective potential lies outside the expected range of validity of the one-loop approximation and, therefore,  it must be rejected as an artifact of the approximation~\cite{Coleman:1973,Jackiw:1974}. 
On the other hand,  the triviality hypothesis  implies that the fluctuation field $\eta$ is a free field with mass $\omega(\phi_0)$. 
As a consequence  the {\it{ exact}} effective potential is:
\begin{equation} 
\label{2.5} 
V_{eff}(\phi_0)  =  \frac{1}{4} \lambda_0 \, \phi_0^4  + \frac{1}{2}  \int \frac{d^3 k}{(2 \pi)^3}  \;  \sqrt{  \vec{k}^2  + \omega^2(\phi_0)  } \; = \;
 \frac{1}{4} \lambda_0 \, \phi_0^4  + \frac{\omega^4(\phi_0)}{64 \pi^2}  \ln \left ( \frac{ \omega^2(\phi_0)}{\Lambda^2} \right )   \;  \; ,
\end{equation}
where $\Lambda$ is an ultraviolet cutoff.  
Moreover, the mechanism of spontaneous symmetry breaking implies that the mass of the fluctuation is related to the scalar condensate as: 
\begin{equation} 
\label{2.6} 
\omega^2(\phi_0) \; = \;  3 \, \tilde{\lambda} \,  \phi_0^2  \;   \; , \; \;  \tilde{\lambda} \; = \; a_1 \, \lambda_0 \; \; ,
\end{equation}
where $a_1$ is some numerical constant.  \\
Now the problem is to see if it exists the continuum limit $\Lambda \rightarrow \infty$.  Obviously, we must have:
\begin{equation} 
\label{2.7} 
\left [   \Lambda \frac{\partial }{\partial  \Lambda} \; + \; \beta(\lambda_0) \, \frac{\partial }{\partial  \lambda_0}   \; + \; \gamma(\lambda_0) \, \phi_0 \,  \frac{\partial }{\partial  \phi_0}
          \right ] V_{eff}(\phi_0)  \; = \; 0   \; .
\end{equation}
Note that in the present case  we cannot use perturbation theory to determine $\beta(\lambda_0)$ and $\gamma(\lambda_0)$. 
Firstly, we note that the  effective potential displays a minimum at:
\begin{equation} 
\label{2.8} 
 3 \tilde{\lambda}  v_0^2 \; = \;   \frac{\Lambda^2}{\sqrt{e}} \; \exp{ [-   \frac{16 \pi^2}{9 \tilde{\lambda}}]} \; ,
\end{equation}
and
\begin{equation} 
\label{2.9} 
V_{eff}(v_0)   \; = \;  - \;  \frac{m_{H_T}^4}{128 \pi^2 } \;  \; , \; \;  m_{H_T}^2 \; = \; \omega^2(v_0) \; .
\end{equation} 
Using Eq.~(\ref{2.7}) at the minimum $v_0$ we get:
\begin{equation} 
\label{2.10} 
\left [   \Lambda \frac{\partial }{\partial  \Lambda} \; + \; \beta(\lambda_0) \, \frac{\partial }{\partial  \lambda_0} 
          \right ] m_{H_T}^2   \; = \; 0   \; ,
\end{equation} 
which in turns gives:
\begin{equation} 
\label{2.11} 
 \beta(\lambda_0)  \;= \; - \, a_1 \; \frac{9}{8 \pi^2} \,  \tilde{\lambda}^2  \; .
\end{equation}
This last equation implies that the theory is free asymptotically for $\Lambda \rightarrow \infty$ consistently  with triviality:
\begin{equation} 
\label{2.12} 
\tilde{\lambda}  \;  \;  \sim \; \;  \frac{16 \pi^2}{9 a_1} \; \frac{1}{\ln(\frac{\Lambda^2}{m_{H_T}^2})}   \; .
\end{equation}
Inserting now Eq.~(\ref{2.11}) into   Eq.~(\ref{2.7}) we obtain:
\begin{equation} 
\label{2.13} 
 \gamma(\lambda_0)  \;= \;  \, a_1^2  \; \frac{9}{16 \pi^2} \,  \tilde{\lambda}  \; .
\end{equation}
Note that this last equation assures that $ \tilde{\lambda} \,  \phi_0^2$ is a renormalization group invariant. 
Rewriting the effective potential as:
\begin{equation} 
\label{2.14} 
V_{eff}(\phi_0)  =  \frac{(3  \tilde{\lambda} \,  \phi_0^2)^2}{64 \pi^2}    \; \left [
  \ln \left ( \frac{ 3  \tilde{\lambda} \,  \phi_0^2}{m_{H_T}^2} \right ) \;  - \; \frac{1}{2}  \right ] \;  \; ,
\end{equation}
we see that $V_{eff}$ is manifestly renormalization group invariant. \\
Let us introduce the renormalized field $\eta_R$ and condensate $\phi_R$. Since the fluctuation $\eta$ is a free field we have $\eta_R = \eta$, namely:
\begin{equation} 
\label{2.15} 
Z_{\eta} \;  =  \; 1 \; . 
\end{equation}
On the other hand, for the scalar condensate according to Eq.~(\ref{2.13}) we have:
\begin{equation} 
\label{2.16} 
\phi_R \; = Z_{\phi}^{-\frac{1}{2}} \;  \phi_0 \; \; \; , \; \; Z_{\phi} \; \sim  \;  \lambda_0^{-1} \sim  \;  \ln(\frac{\Lambda}{m_H})  \;  . 
\end{equation}
As a consequence we get that the physical mass $m_{H_T}$ is {\it finitely} related to the renormalized vacuum expectation scalar field value $v$:
\begin{equation} 
\label{2.17} 
m_{H_T} \;  =  \; \xi \; v  \; . 
\end{equation}
It should be clear that the physical mass $m_{H_T}$ is an arbitrary parameter of the theory (dimensional transmutation). 
On the other hand the parameter $\xi$ being a pure number can be determined in the non perturbative lattice approach. 
Remarkably,  extensive numerical simulations  showed that   the physical Higgs boson mass $m_{H_T}$
is finitely related to the renormalized vacuum expectation value $v$. Moreover, the extrapolation to the continuum limit of the ratio 
$m_{H_T}/v$ leads to  the intriguing  relation~\cite{Cea:2012}: 
\begin{equation}
\label{2.18}
\xi \; \simeq \;  \pi  \;  ,
\end{equation}
pointing to  a rather massive  $H_T$ boson, namely $m_{H_T} \; \simeq \; 750$ GeV. 
\section{The $H_T$ Higgs boson}
\label{s-3}
In the previous Section we presented simple arguments to illustrate  how  the  spontaneous symmetry breaking  mechanism
can be implemented also  for  real scalar fields without self-interaction. One could object that our treatment  could not be relevant 
for the physical Higgs boson, for the scalar theory relevant for the Standard Model is
the O(4)-symmetric self-interacting theory. However, the Higgs mechanism eliminates three scalar fields leaving as physical Higgs field  the radial excitation whose dynamics  is described by  the one-component self-interacting scalar field theory. Therefore, we are confident that our 
theoretical arguments  can be reliably extended  to the Standard Model Higgs boson. \\
In order to determine the phenomenological signatures of the massive $H_T$ Higgs boson we need to take care of the couplings with 
the gauge and fermion fields of the Standard Model~\footnote{A preliminary  account on this matter has been presented in the second part
of Ref.~\cite{Cea:2016}.}. Actually, the coupling of the Higgs field to the gauge vector bosons is fixed by
 the gauge symmetries. Therefore the coupling of
 the $H_T$ Higgs boson to the  gauge vector bosons  is the same as in perturbation theory notwithstanding the
 non perturbative Higgs condensation driving the spontaneous breaking of the gauge symmetries. 
Given the rather large mass of the $H_T$ Higgs boson, the main decay modes are the decays into two massive
vector bosons (see, e.g., Refs.~\cite{Gunion:1990,Djouadi:2008}):
\begin{equation}
\label{3.1}
\Gamma( H_T \; \rightarrow \; W^+ \, W^-)  \; \simeq  \;  \frac{G_F \, m^3_{H_T}}{8 \pi \sqrt{2}} \;
 \sqrt{1 - \frac{4 m^2_W}{m^2_{H_T}}} \;  \bigg ( 1 - 4 \,  \frac{m^2_W}{m^2_{H_T}} + 12 \, \frac{ m^4_W}{m^4_{H_T}}
 \bigg ) \; 
\end{equation}
and
\begin{equation}
\label{3.2}
\Gamma( H_T \;  \rightarrow \;  Z^0 \, Z^0) \; \simeq \;   \frac{G_F \, m^3_{H_T}}{16 \pi \sqrt{2}} \;
 \sqrt{1 - \frac{4 m^2_Z}{m^2_{H_T}}} \; \bigg ( 1 - 4 \,  \frac{m^2_Z}{m^2_{H_T}} + 12 \, \frac{ m^4_Z}{m^4_{H_T}}
 \bigg )  \; . 
\end{equation}
 On the other hand, it is known that for heavy Higgs the radiative corrections to the decay widths can be safely 
 neglected~\cite{Fleischer:1981,Fleischer:1983,Marciano:1988}. \\
 The couplings of the $H_T$ Higgs boson to the fermions are given by the Yukawa couplings $\lambda_f$. Unfortunately,
 there are not reliable lattice non-perturbative simulations on the continuum limit of the Yukawa couplings. If we follow the
 perturbative approximation, then the fermion Yukawa couplings turn out to be proportional to the fermion mass,
 $\lambda_f = \sqrt{2} \, m_f/v$. In that case, for heavy Higgs the only relevant fermion coupling is the top Yukawa coupling 
 $\lambda_t$. 
 On the other hand, we cannot exclude that the couplings of the physical Higgs field  to the fermions
 could be very different from  perturbation theory. Indeed, the non-trivial rescaling of the Higgs condensate suggests that, 
if the fermions acquires a finite mass through the Yukawa couplings,
then the coupling of the physical Higgs field to the fermions could  be strongly suppressed. 
In this case, the whole issue of  generation of fermion masses through the Yukawa couplings should  be reconsidered. 
Therefore, to parametrize  our ignorance on the Yukawa couplings of the $H_T$ Higgs boson we introduce the parameter:
\begin{equation}
\label{3.3}
\kappa \; = \; \lambda_t^2 \; \frac{v^2}{2 \, m_t^2} \; .
\end{equation}
Obviously, in perturbation theory we have $\kappa = 1$. Nevertheless, we shall also consider the case  $\kappa \simeq 0$
corresponding to strongly suppressed fermion  Yukawa  couplings. \\
The width for the decays of the $H_T$ boson into a $t \bar{t}$ pairs is easily found~\cite{Gunion:1990,Djouadi:2008}:
\begin{equation}
\label{3.4}
\Gamma( H_T \rightarrow \; t \, \bar{t}) \; \simeq \; \kappa \,   \frac{3 \, G_F \, m_{H_T} m^2_t}{4 \pi \, \sqrt{2}} \;
\bigg ( 1 - 4 \,  \frac{m^2_t}{m^2_{H_T}}  \bigg )^{\frac{3}{2}}  \; . 
\end{equation}
So that, to a good approximation, the Higgs total width is given by:
\begin{equation}
\label{3.5}
\Gamma_{H_T}  \; \simeq \; \Gamma( H_T \rightarrow W^+ \, W^-)  \; + \; \Gamma( H_T \rightarrow Z^0 \, Z^0)  \; + \;
 \Gamma( H_T \rightarrow t \, \bar{t})  \; .
\end{equation}
Our previous equations show  that, in the high mass region $m_{H_T} \gtrsim 400 \,  GeV$  the  total width depends
 strongly on the Higgs mass.
The main difficulty in the experimental identification of a very heavy  Higgs   resides in the  
large width which makes almost impossible to observe a mass peak.  In fact, the expected  mass spectrum of our heavy Higgs 
 should be proportional to the Lorentzian distribution.  For a resonance with mass $M$ and total width $\Gamma$
 the Lorentzian distribution is given by:
\begin{equation}
\label{3.6}
L (E) \;  \sim  \;  \; \frac{\Gamma}{(E \; - \; M^2)^2 \; + \;  \Gamma^2/4}  \;  .
\end{equation}
Note that  Eq.~(\ref{3.6}) is the simplest distribution consistent with the Heisenberg uncertainty principle and the finite lifetime
$\tau \simeq 1/\Gamma$.  We obtain, therefore, the following Lorentzian  distribution:
\begin{equation}
\label{3.7}
L_ {H_T} (E) \;  \simeq  \;  \; \frac{1}{ 1.0325 \;  \pi} \; \frac{\frac{\Gamma_{H_T}(E)}{2}}{\big (E \; - \; 750 \; GeV \big )^2 \; + 
\;  \big ( \frac{\Gamma_{H_T}(E)}{2} \big )^2} \;  \;   \;  \; , 
\end{equation}
where $\Gamma_{H_T} ( E)$ is given by Eq.~(\ref{3.5}), and the normalization is  such that:
\begin{equation}
\label{3.8}
\int^{\infty}_{0} \; L_ {H_T} (E)  \;  dE \; \; = \; \; 1 \; \; .
\end{equation}
Note that, in the limit $\Gamma_{H_T} \; \rightarrow \; 0$,  $L_ {H_T} (E)$ reduces to $\delta(E \; - 750 \; GeV)$.  \\
To evaluate the Higgs event production at LHC we need the inclusive Higgs production cross section. As in perturbation
theory, for large Higgs masses the main production processes are by vector-boson fusion and gluon-gluon fusion. 
In fact, the $H_T$ Higgs production cross section by vector-boson fusion is the same as in the perturbative Standard Model calculations.
Moreover, for Higgs mass in the range $700 - 800 \; GeV$ the main production mechanism at LHC is 
expected to be  by the gluon fusion mechanism.  The gluon coupling to the Higgs boson in the Standard Model is 
mediated by triangular loops of top and bottom quarks.  Since in perturbation theory the Yukawa couplings 
of the Higgs particle to heavy quarks grows with quark mass, thus balancing the decrease of the triangle amplitude, 
the effective gluon coupling  approaches a non-zero value for large loop-quark masses. This means that for
heavy Higgs the gluon fusion inclusive cross section is almost completely  determined by the top quarks, even though
it is interesting to stress that  for large Higgs masses the vector-boson fusion mechanism becomes competitive to 
the gluon fusion Higgs production.
According to our approximations the total inclusive cross section for the production of the $H_T$ Higgs boson
can be written as:
\begin{equation}
\label{3.9}
\sigma(p \; p \;  \rightarrow \; H_T \; + \; X) \; \simeq \;    \sigma_{VV}(p \; p \;  \rightarrow \; H_T \; + \; X)
\; + \; \kappa \; \sigma_{gg}(p \; p \;  \rightarrow \; H_T \; + \; X)  \; ,
\end{equation}
where  $\sigma_{VV}$ and   $\sigma_{gg}$ are the vector-boson fusion and gluon-gluon fusion inclusive cross
sections respectively. \\
\begin{figure}
\includegraphics[width=0.5\textwidth,clip]{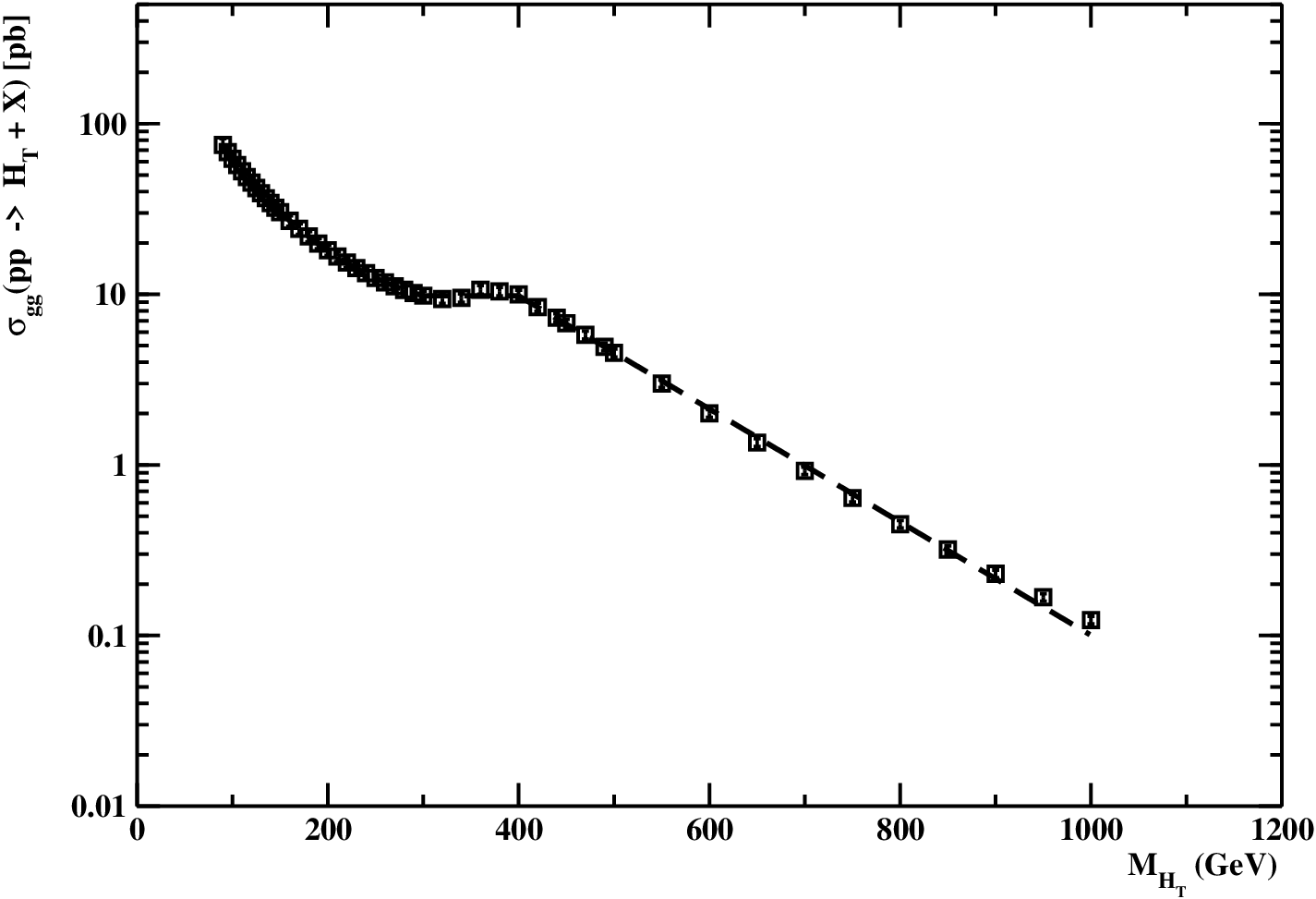}
\hspace{0.2 cm}
\includegraphics[width=0.5\textwidth,clip]{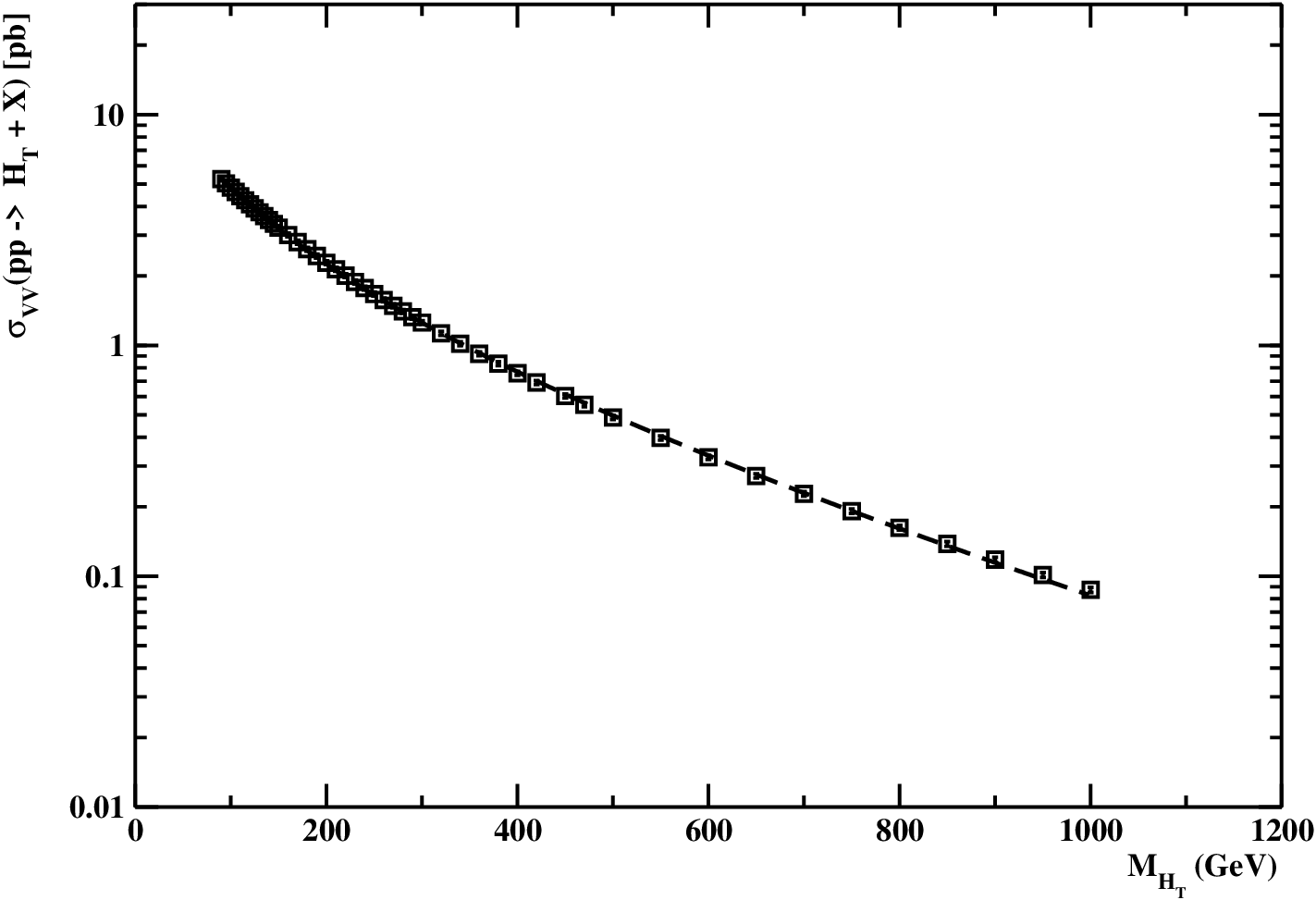}
\caption{\label{Fig1} The inclusive Higgs production gluon-gluon fusion (left panel) and vector-boson fusion (right panel) cross sections 
at $\sqrt{s} =  13 \, TeV$. The data have been taken from  Ref.~\cite{Hcross13Tev}. The dashed lines are the fits to the data
with  our parametrization Eqs.~(\ref{3.10}) and (\ref{3.12}).}
\end{figure}
In Ref.~\cite{Hcross13Tev} it is presented the calculations of the cross sections computed at next-to-next-to-leading 
and next-to-leading order  for    heavy Higgs boson with Standard Model-like coupling
 at $\sqrt{s} = 13 \, TeV$. In Fig.~\ref{Fig1} we show 
the dependence on the Higgs mass of the perturbative  gluon-gluon (left panel) and vector-boson fusion (right panel) 
cross sections at $\sqrt{s} =13 \,  TeV$  as reported in  Ref.~\cite{Hcross13Tev}.   
As concern the gluon-gluon fusion cross section  we found that  this cross section can be
 parametrized as:
\begin{equation}
\label{3.10}
 \sigma_{gg}(p \; p \;  \rightarrow \; H_T \; + \; X)  \; \simeq \;  
 \left\{ \begin{array}{ll}
 \;  \left (  \frac{ a_1}{ M_{H_T}} 
 \; + \; a_2 \; M_{H_T}^3  \right )  \;  \exp (-  a_3 M_{H_T})  \; \; &  M_{H_T}  \; \leq \; 300  \; GeV 
  \\
 \; \; \;  \; a_4  \;  & 300 \; GeV    \leq  M_{H_T}   \leq  400 \; GeV
  \\
 \; \;  \; \;a_4 \;  \exp \big [ - a_5 ( M_{H_T} - 400 \; GeV) \big ]  \; \; &  400  \; GeV \; \leq \; M_{H_T}
\end{array}
    \right.
\end{equation}
where $M_{H_T} $  is expressed in  GeV.  In fact we fitted Eq.~(\ref{3.10}) to the data (see the dashed line in 
Fig.~\ref{Fig1}, left panel) and obtained:
\begin{eqnarray}
\label{3.11}
a_1 \simeq 1.24 \, 10^4 \; pb \, GeV \;  \; , \; \;  a_2 \simeq 1.49 \, 10^{-6} \; pb \, GeV^{-3} \; , \;  
\nonumber \\
a_3 \simeq 7.06 \, 10^{-3} \;  GeV^{-1} \; , \;  \; a_4 \simeq 9.80 \;\, pb  \; , \hspace{2.2 cm}
\\ \nonumber
a_5 \simeq 7.63 \, 10^{-3}  \; GeV^{-1}  \; . \hspace{5.05 cm}
\end{eqnarray}
Likewise,  we parametrized the  dependence of the vector-boson fusion cross section as:
\begin{equation}
\label{3.12}
 \sigma_{VV}(p \; p \;  \rightarrow \; H_T \; + \; X) \; \simeq \;    \bigg ( b_1 \; + \;  \frac{ b_2}{ M_{H_T}} 
 \; + \; \frac{b_3}{ M_{H_T}^2}  \bigg )  \;  \exp (-  b_4 \;  M_{H_T} )   \; ,
\end{equation}
and obtained for the best fit (see the dashed line in  Fig.~\ref{Fig1}, right panel) :
\begin{eqnarray}
\label{3.13}
b_1 \simeq - 2.69 \, 10^{-6}  \; pb   \;  \; , \; \;  b_2 \simeq 8.08 \, 10^{2} \; pb \, GeV \; , \hspace{1.15 cm}
 \nonumber \\
b_3 \simeq - 1.98 \, 10^{4}  \; pb \,  GeV^{2}  \; \;  , \;  b_4 \simeq  2.26 \, 10^{-3} \; GeV^{-1} \; .  \; \; \,
\end{eqnarray}
\section{$H_T$ Decay Channels }
\label{s-4}
We have seen that the main decays of the heavy $H_T$ Higgs boson are the decays into two massive vector boson
and $t \bar{t}$ pairs, if $\kappa = 1$. Since the search for the decays of the $H_T$ Higgs boson into
pairs of top quarks is very complicated because of the large QCD background, we will focus on the decays
into two massive vector bosons. It is worthwhile to stress that the $H_T$ decay to $\gamma \gamma$
do not proceed directly, but, to a fair good approximation, via longitudinal $W$ virtual states. Therefore the relevant
branching ratio turns out be strongly suppressed, i.e. $Br( H_T \rightarrow \gamma \gamma) \sim 10^{-6}$~\cite{Ellis:1976,Shifman:1979}. 
\\
To compare the invariant mass spectrum of our $H_T$ Higgs with the experimental data, we observe that:
\begin{equation}
\label{4.1}
 N_{H_T} (E_{1},E_{2} )  \; \simeq \; {\cal{L}} \;  \int^{E_2}_{E_1} \; Br(E) \;  \varepsilon(E) \; \sigma(p \; p \;  
 \rightarrow \; H_T \; + \; X)  \; L_ {H_T} (E)  \;  dE  
 \;  \; , 
\end{equation}
where  $N_ {H_T}$ is the number of Higgs events in the energy interval $E_1,E_2$, corresponding to an integrated luminosity 
${\cal{L}}$, in the given channel with branching ratio $Br(E)$.   The parameter $ \varepsilon(E)$  accounts for  the efficiency 
of trigger, acceptance of the detectors, the kinematic selections, and so on.  Thus, in general
 $ \varepsilon(E)$ depends on the energy, the selected channel and the detector.   
 For illustrative purposes, we consider the decay  channels  $H  \rightarrow WW  \rightarrow fermions$ and 
 $H  \rightarrow ZZ   \rightarrow fermions$. Thus, for the branching ratios Br(E)  we can write:
\begin{equation}
\label{4.2}
  \begin{array}{ll}
Br(H_T  \rightarrow WW  \rightarrow fermions ) \;  \simeq  \; Br(H_T  \rightarrow WW  ) \;
\times  \; Br( WW  \rightarrow fermions ) 
  \\
Br(H_T  \rightarrow ZZ   \rightarrow fermions )  \;  \simeq  \; Br(H_T  \rightarrow ZZ  )  \;
\times  \; Br( ZZ \rightarrow fermions ) \; ,
\end{array}
\end{equation}
where
\begin{equation}
\label{4.3}
Br(H_T  \rightarrow WW  ) \; = \; \frac{ \Gamma( H_T \rightarrow WW)}{ \Gamma_{H_T}}   \; \; , \; \; 
Br(H_T  \rightarrow ZZ  ) \; = \; \frac{ \Gamma( H_T \rightarrow ZZ)}{ \Gamma_{H_T}}   \; ,
\end{equation}
while the branching ratios for the decays of $W$ and $Z$ bosons into fermions are given by the Standard Model 
values~\cite{Olive:2014}. \\
\begin{figure}
\includegraphics[width=0.5\textwidth,clip]{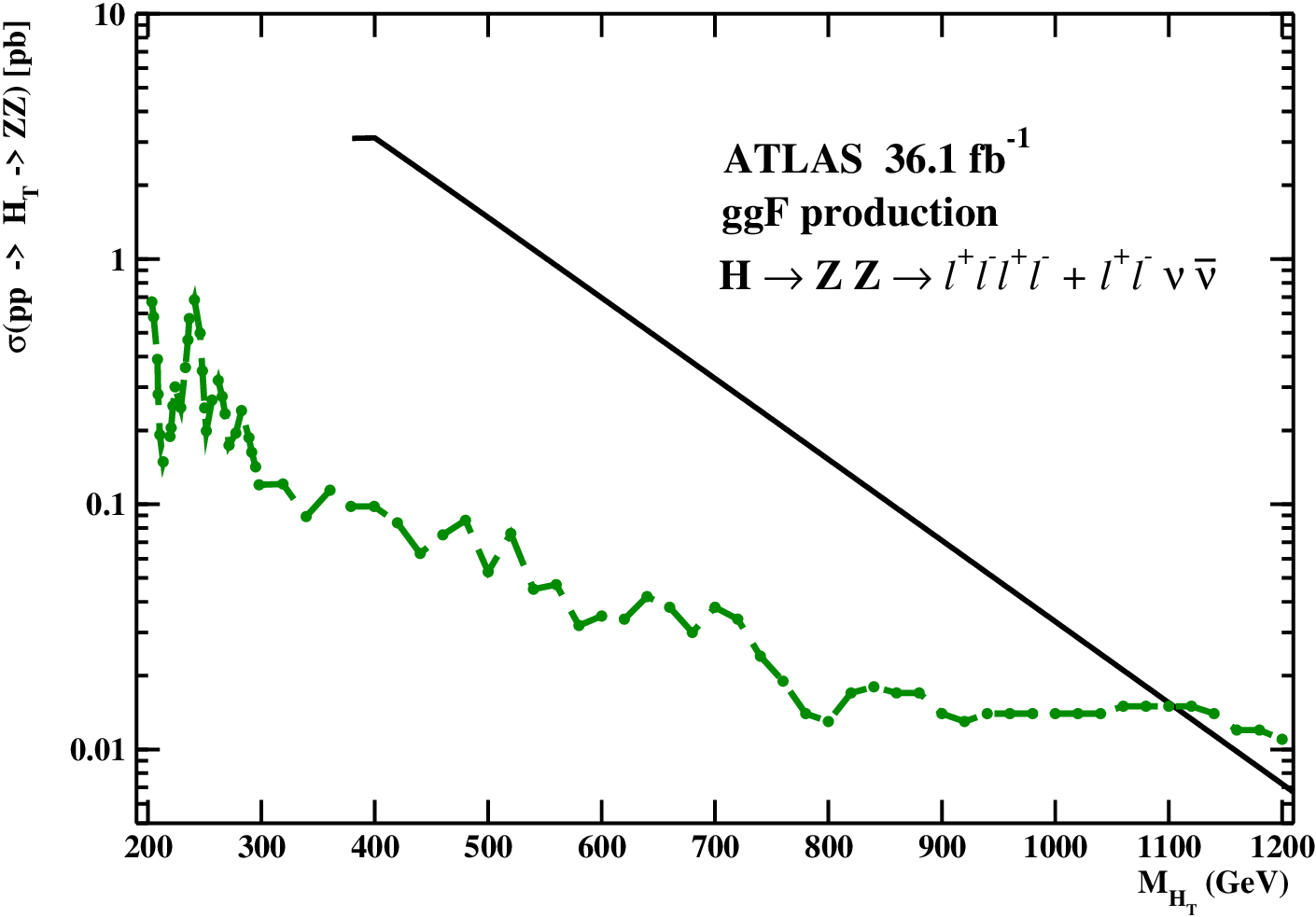}
\hspace{0.2 cm}
\includegraphics[width=0.5\textwidth,clip]{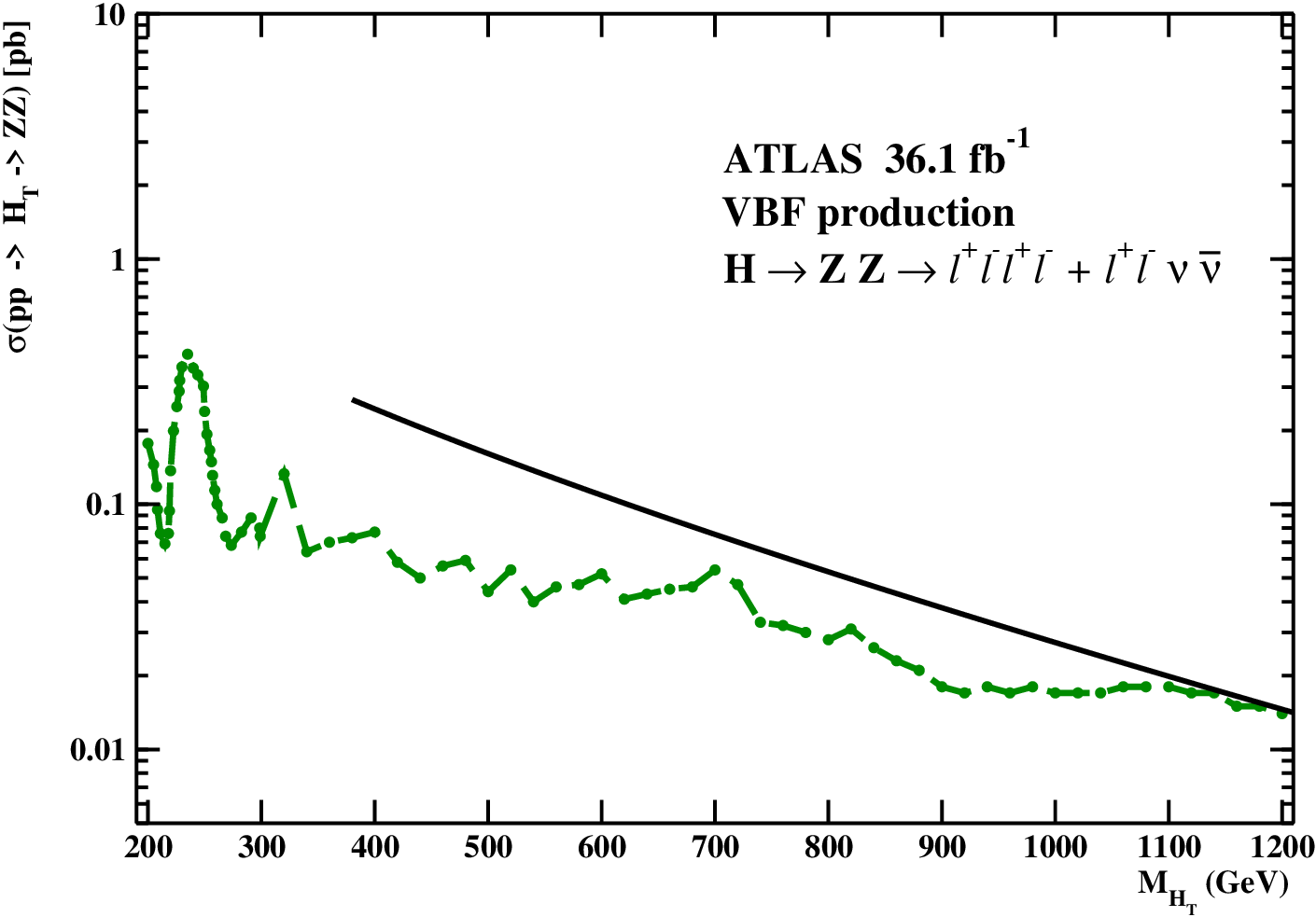}
\caption{\label{Fig2}(color online) Observed  limits on the cross-section times branching ratio to ZZ final states
for a narrow-width heavy scalar resonance  as a function of its mass. The limits corresponds to the decays of the two Z vector bosons 
 into  $\ell^+ \ell^- \ell^+ \ell^-$  or  $\ell^+ \ell^-  \nu \bar{\nu}$, where $\ell$ is an electron or muon. 
 The limits correspond to gluon-gluon fusion  (left panel)  and  vector-boson fusion  (right  panel)   production mechanisms. 
 The data (dashed green lines) have been obtained from Fig.~6 in   Ref.~\cite{ATLAS:2017c}.  The continuum black lines are our theoretical estimates of the cross-section times branching ratio  for gluon-gluon fusion processes, Eqs.~(\ref{3.10}) and (\ref{3.11}), and vector-boson fusion processes, Eq.~(\ref{3.12}) and (\ref{3.13}).}
\end{figure}
In Fig.~{\ref{Fig2}  we show the observed limits for a new neutral spin-zero resonance decaying to two Z bosons.
The limits on the inclusive production cross section times the relevant branching ratio (the dashed lines in Fig.~{\ref{Fig2}) 
have been obtained by the ATLAS Collaboration from the analysis of the data  collected  during the LHC Run 2 
with an integrated luminosity of $36.1 \, fb^{-1}$~\cite{ATLAS:2017c} by using the decay channels where the pair of Z bosons decay
 leptonically to  $\ell^+ \ell^- \ell^+ \ell^-$   or  $\ell^+ \ell^-  \nu \bar{\nu}$ final states, $\ell$ being  either
  an electron or a muon. 
 The displayed limits correspond to gluon-gluon fusion  (left panel)  and  vector-boson fusion  (right  panel)   production mechanisms
 by assuming   a narrow width  Higgs-like scalar resonance.  This allows us to  directly compare the data
to our theoretical estimates of the inclusive production cross section, Eqs.~(\ref{3.10}) and (\ref{3.12})
 times the branching ratios as given by Eq.~(\ref{4.3}).
In fact, in Fig.~{\ref{Fig2} the continuum lines are our theoretical  cross-section times branching ratio 
for gluon-gluon fusion processes, Eqs.~(\ref{3.10}) and (\ref{3.11}), and vector-boson fusion processes, Eq.~(\ref{3.12})
and (\ref{3.13}). Of course, for our purposes  the relevant region is the high-mass  region near  $750 \, GeV$.  
From  Fig.~{\ref{Fig2}  we see that the experimental data display some wiggles but no significant excess with respect to expected
background signal is still observed. Nevertheless, it is intriguing to observe that the experimental limits  for the vector-boson
inclusive production mechanism display a bump in the invariant mass region around  $700 \, GeV$ that, albeit not yet statistically significant, 
may hint at the presence of a signal. Note that the bump near   $700 \, GeV$ is mainly due to the four-lepton final states.
Moreover, taking into account the fact that for a broad scalar resonance the experimental limits are expected to become
somewhat weaker, we see that our theoretical vector-boson fusion inclusive cross section is in fair agreement with the 
experimental observations. Indeed, we shall show in the following Section that the observed invariant mass distribution
for the  $H_{T} \rightarrow ZZ \rightarrow \ell  \ell  \ell  \ell $ decay channels is in quite good agreement with
our theoretical expectations. On the other hand, looking at Fig.~{\ref{Fig2}, left panel, it is evident that  the experimental data 
are not consistent with our theoretical estimate of the gluon-gluon fusion inclusive
production cross section.  Therefore,  the preliminary data from the LHC Run 2 are clearly indicating that $\kappa \simeq 0$, namely
the couplings of our $H_T$ Higgs boson  to fermions is strongly suppressed. We see, thus, that  the inclusive production of the 
 $H_T$ Higgs boson in proton-proton collisions  is due mainly to vector boson fusion processes. This means that, at moment, there is not
 enough sensitivity in the decay channels where one or both the vector bosons decay into hadrons since
the expected signal is within the statistical uncertainties in the Monte Carlo evaluations of the huge QCD background.
\subsection{The golden channel}
\label{s-4-1}
In the present Section we attempt a direct comparison of our theoretical expectations with the available experimental data
from LHC Run 2 in the so-called golden channel corresponding to the  decays $H_T  \rightarrow ZZ  \rightarrow \ell \ell  \ell \ell $.
Indeed, the four-lepton channel, albeit rare, has the clearest and cleanest signature of all the possible 
Higgs boson decay modes due to the small background contamination. 
\begin{figure}
\includegraphics[width=0.5\textwidth,clip]{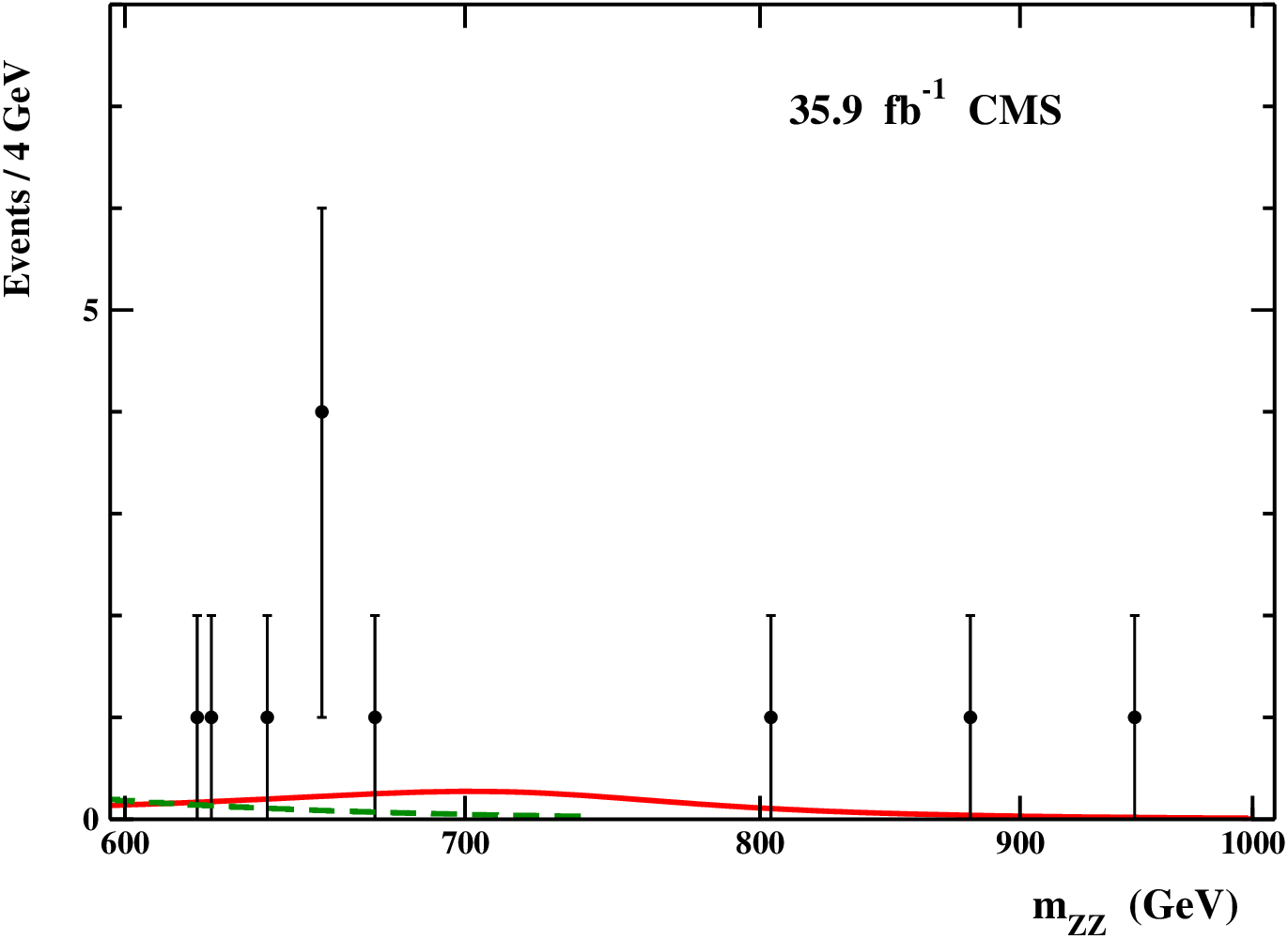}
\hspace{0.2 cm}
\includegraphics[width=0.5\textwidth,clip]{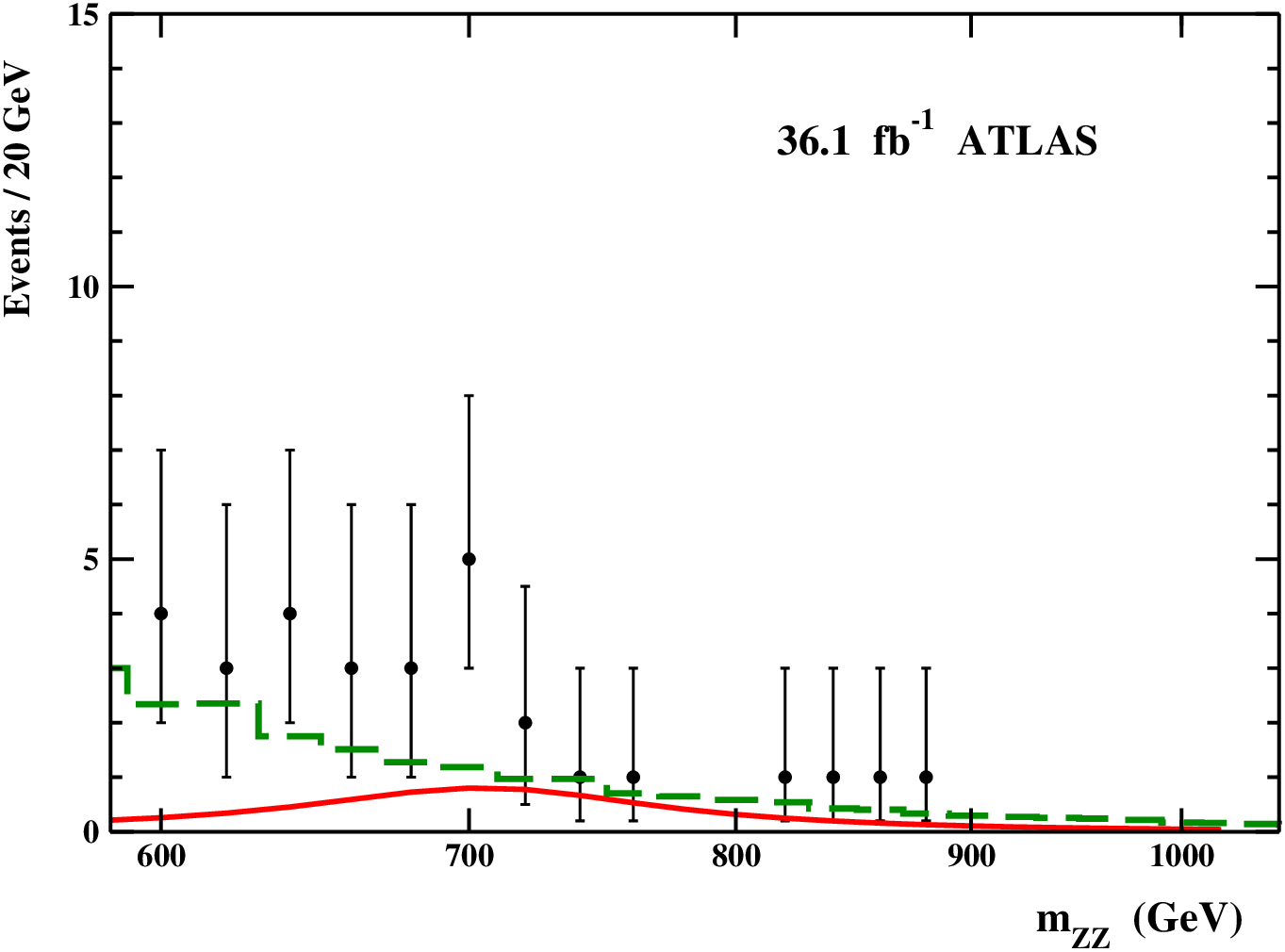}
\caption{\label{Fig3} (color online) Comparison to the LHC data of the distribution of the invariant mass $m_{Z Z}$ for the  process
 $H_T \; \rightarrow ZZ \; \rightarrow \ell \ell \ell \ell$   ($\ell = e, \mu$)  in the high-mass region   $m_{Z Z}  \gtrsim 600 \,  GeV$.
The CMS data (left panel)  have been obtained  from Fig.~3, left panel, in Ref.~\cite{CMS:2017z} 
 and correspond   to  an  integrated luminosity of  ${\cal{L}} = 35.9 \,  fb^{-1}$. The ATLAS data (right panel),
 corresponding  to  an  integrated luminosity of  ${\cal{L}} = 36.1 \,  fb^{-1}$,  have
 been obtained  from Fig.~4, left panel, in Ref.~\cite{ATLAS:2017c}. 
The dashed (green) lines are our estimate of the background, the continuum (red) lines are the  signal histogram assuming 
$\varepsilon(E) \simeq  0.80$ and $\kappa \simeq 0$. }
\end{figure}
In Fig.~\ref{Fig3} we show  the invariant mass distribution for the golden channel obtained  from the CMS experiment with
an integrated luminosity of   $35.9 \, fb^{-1}$~\cite{CMS:2017z} (left panel) and the ATLAS experiment with
an integrated luminosity of   $36.1 \, fb^{-1}$~\cite{ATLAS:2017c} (right panel).  From our estimate of the background 
(dashed lines in Fig.~\ref{Fig3}) we see that, indeed, in the high invariant mass region   $m_{ZZ}  \gtrsim 650 \,  GeV$, 
the background is strongly suppressed.  To compare  with our theoretical expectations, we displayed 
in  Fig.~\ref{Fig3}  the distribution of the invariant mass for the $H_T$ Higgs boson candidates corresponding to the golden channel. 
 The distributions have been obtained using Eq.~(\ref{4.1})  with $\kappa = 0$ and $\varepsilon(E) \simeq  0.80$  to take care of the fact that the detectors do not cover the full phase space. We also assumed a slightly smaller value for the heavy Higgs boson central mass, namely
 $m_{H_T} \simeq 730 \, GeV$ that, however, is within the  statistical uncertainties of the lattice determination~\cite{Cea:2012}. 
\\
Due to the rather low integrated luminosity, we see that the signal manifest itself
as a broad plateaux in the   invariant mass interval   $650 \, GeV \lesssim m_{ZZ}   \lesssim  1000 \,  GeV$
over  a smoothly decreasing background. Actually, in this region it seems that  there is a moderate excess of signal
over the expected background that seems to  compare quite well with our theoretical prediction. 
To be qualitative, we may estimate  the total number of events in the invariant mass interval  
$650 \, GeV \lesssim m_{ZZ}   \lesssim  1000 \,  GeV$ and compare with our theoretical expectations.
We find:
\begin{equation}
\label{4.4}
N_{obs} = 8.0^{+5.00}_{-2.83}   \; ,  \;
 N_{back} = 1.73     \; ,  \;
N^{obs}_{sign} = 6.27^{+5.00}_{-2.83}     \; , \; 
N^{th}_{sign} = 5.93   \; \;  \; \;   CMS \; \; \; 
\end{equation}
\begin{equation}
\label{4.5}
N_{obs} = 23.0^{+8.14 }_{-4.70}   \; , \; 
 N_{back} = 11.0     \; , \; 
N^{obs}_{sign} = 12.00^{+8.14 }_{-4.70}      \; , \; 
N^{th}_{sign} =   5.96   \; \;  \;   ATLAS \;   \;  
\end{equation}
\begin{figure}
\centering
\includegraphics[width=0.65\textwidth,clip]{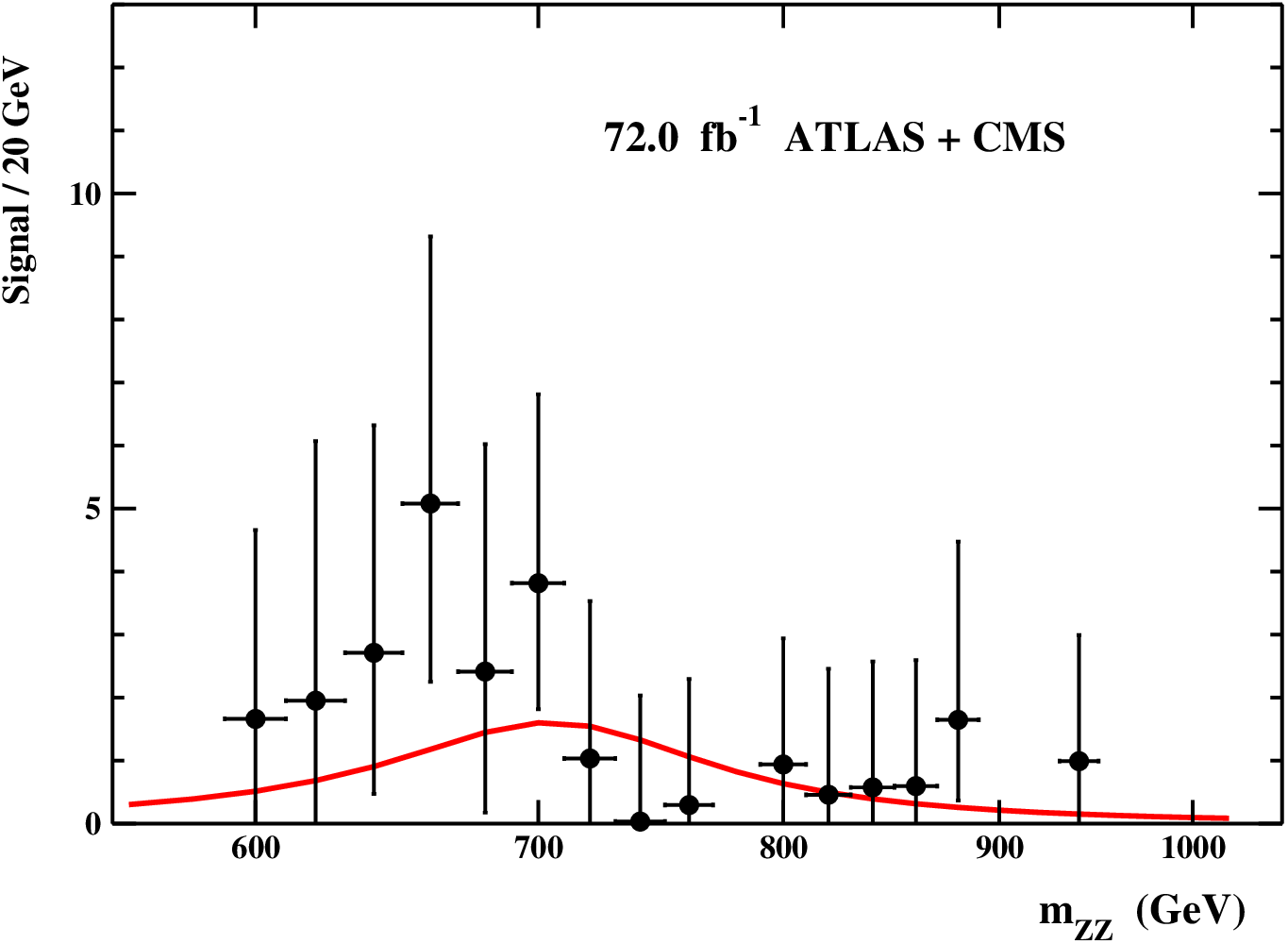}
\caption{\label{Fig4} (color online) Comparison to the LHC data of the distribution of the invariant mass $m_{Z Z}$
in the high-mass region   $m_{Z Z}  \gtrsim 600 \,  GeV$  for the  process
 $H_T \; \rightarrow ZZ \; \rightarrow \ell \ell \ell \ell$   ($\ell = e, \mu$).
The signal distribution  has been obtained from the combination of  the ATLAS and CMS event distributions  by
subtracting the relevant background.  The continuum (red) line is the  expected signal histogram assuming 
$\varepsilon(E) \simeq  0.80$ and $\kappa \simeq 0$.}
\end{figure}
 where,  to be conservative, the quoted errors have been obtained by adding in quadrature the experimental errors.
Remarkably, both ATLAS and CMS distributions  display  an excess over the expected background with a statistical significance of 
more than two standard deviations. Moreover, the excesses  interpreted as a signal seem to be in fair agreement  with our theoretical expectations.
To better appreciate this point, in Fig.~\ref{Fig4} we show the signal-distribution of the invariant mass  $m_{Z Z}$
in the high-mass region   $m_{Z Z}  \gtrsim 600 \,  GeV$. The signal has been obtained by combining the ATLAS and CMS binned events
and subtracting the relevant background. It is remarkable that the signal distribution displays a broad peak structure around
 $m_{Z Z}  \sim 700 \,  GeV$ with a statistical significance well above three standard deviations. Moreover, we see that our
 theoretical signal distribution (continuum line in  Fig.~\ref{Fig4}) is in reasonable agreements with the experimental data.
 Therefore, we may conclude that our proposal for the heavy  $H_T$ Higgs boson  is finding in the golden channel the first confirmation,
 even though  we cannot yet completely exclude the compatibility of the data  with  the background-only hypothesis. 
\section{Conclusion}
\label{s-5}
It is widely believed that the new LHC resonance at $125 \, GeV$  is the Standard Model Higgs boson. 
However, stemming from the known triviality problem, i.e. vanishing self-coupling, that affects self-interacting scalar quantum fields
in four space-time dimensions, we evidenced that the Higgs boson condensation triggering the spontaneous
breaking of the local gauge symmetries needs to be dealt with non perturbatively. It is worthwhile to notice that in this case,
from one hand there is no stability problem for the condensate ground state, on the other hand the Higgs mass is finitely related to the vacuum 
expectation value of the quantum scalar field and, in principle, it can be evaluated from first principles.
In fact, precise non-perturbative numerical simulations~\cite{Cea:2012} gave for the $H_T$ Higgs boson mass 
$m_{H_T} = 754 \pm 20 \, GeV$ leading to a rather heavy Higgs boson.  In this paper we elaborated some phenomenological
aspects of the heavy Higgs boson scenario. We have critically discussed the couplings of the $H_T$
Higgs boson to the massive vector bosons and to fermions. We have also estimated the expected production mechanism
and the main decay modes. Comparing with the available LHC Run 2 data we concluded that the coupling of the $H_T$ Higgs
boson to fermions were strongly suppressed.
Finally, we  compared our proposal with the recent results in the golden channel from both
ATLAS and CMS Collaborations. We found that the available experimental observations were consistent with our scenario. 
We are confident that forthcoming data from LHC Run 2 will add further support to the heavy Higgs proposal. Indeed, assuming
a real signal,  by the end of the LHC Run 2  it is expected that both ATLAS and CMS experiments will  reach in the golden channel
a statistical significance of about five standard deviations.

\end{document}